\documentstyle[12pt,epsf]{article}
\input epsf

\def\li2{{\rm Li}_2}

\def\roughly#1{\,\,\raise.3ex\hbox{$#1$\kern-.75em\lower1ex\hbox{$\sim$}}\,\,}

\def\beq{\begin{equation}}
\def\eeq{\end{equation}}
\def\bea{\begin{eqnarray}}
\def\eea{\end{eqnarray}}

\def\det{{\rm Det}}
\def\tr{{\rm Tr}}

\def\prl#1#2#3{Phys. Rev. Lett. {\bf #1} (#2) #3}
\def\mpl#1#2#3{Mod. Phys. Lett. {\bf A#1} (#2) #3}

\def\gsim{{~\raise.15em\hbox{$>$}\kern-.85em
          \lower.35em\hbox{$\sim$}~}}
\def\lsim{{~\raise.15em\hbox{$<$}\kern-.85em
          \lower.35em\hbox{$\sim$}~}}

\begin{document}
\begin{titlepage}
\begin{center}
May 1997  \hfill    CERN-TH/97-85\\
               \hfill    hep-ph/9705348
\vskip .2in
{\large \bf 
Dynamical Inflation and Unification
Scale\\ on \\Quantum Moduli Spaces 
}
\vskip .3in

\vskip .3in
S. Dimopoulos$^{(a,b)}$, G. Dvali$^{(a)}$ and R. Rattazzi$^{(a)}$ 
 \\[.03in]

{$^{(a)}$\em Theory Division, CERN\\
     CH-1211 Geneva 23, Switzerland}

{$^{(b)}$ \em Physics Department, Stanford University\\ 
Stanford, CA 94305, USA}

\end{center}
\vskip .2in
\begin{abstract}
\medskip

We show that simple strongly coupled supersymmetric gauge theories   
with   quantum moduli spaces can naturally lead to  hybrid inflation. 
These   theories  
contain no input dimensionful or small parameters.
The effective superpotential is linear in the inflaton field; this
 ensures that supergravity corrections do not spoil the slow roll  
conditions for inflation. 
We construct a simple theory in which the classical moduli  
space exhibits neither GUT-symmetry-breaking nor inflation  
whereas its quantum modification exhibits both. As a result,  
the dynamical origin and scales of inflation and grand  
unification  coincide.

\end{abstract}
\end{titlepage}

 The inflationary solution \cite{inf} to the standard big bang  
problems assumes that the Universe
has gone through a period of de Sitter phase with  an exponentiallly 
growing scale factor. The temporarily nonzero cosmological  
``constant''
is usually associated to the potential energy of a slowly rolling  
scalar field,
the  inflaton. In order to have successful inflation  the
slow-roll
conditions must be satisfied. These conditions imply that the  
inflaton potential is very
flat on  mass scales of the order of the  Hubble  
parameter. Inflation ends whenever this is not the case.
 Most inflationary scenarios exhibit two aesthetic
problems: {\it i}) the desired
potential is arranged by  invoking an extremely small coupling  
constants
and/or an input ``hard'' mass scale; {\it ii}) the
sector that drives inflation is not usually motivated by particle  
physics considerations.
In this paper we want to propose a class of theories where both problems
are addressed.
We will attempt to naturally reconcile the breaking of the grand  
unified
gauge symmetry with a successful inflationary scenario, within a  
single
dynamical mechanism, with no input mass parameters and/or very small
coupling constants.
Both the value of the inflationary potential and the GUT scale is set
by a single dynamically generated mass parameter, which
 COBE  measurements indicate to be around the scale 
of supersymmetric GUTs. Our inflationary scenario can be viewed as a
natural realization of ``hybrid'' inflation \cite{hybrid}.
In order to give a general idea, let us consider the
``idealized'' case of an exactly
flat potential for the inflaton field $S$
\begin{equation}
     V(S) = \Lambda^4
\end{equation}
In the (globally) supersymmetric context this simply corresponds to  
a linear superpotential
\begin{equation}
  W(S) = S\Lambda^2
\end{equation}
This exhibits a continuous $R$-symmetry and breaks supersymmetry  
through the
expectation value $F_S = \partial _S W=\Lambda^2$.
Of course, in order to have a realistic
situation one needs in addition: $i)$ a force that drives the  
inflaton field;
$ii)$ a trigger which sooner or later  switches off the  
cosmological constant $\Lambda^4$ thus ending inflation.
Both these conditions are automatically satisfied provided there are
particles that  get masses by  coupling to $S$. The  
simplest way is to have\cite{copeland},\cite{dss}
\begin{equation}
  W_{tree} =  \Lambda^2S + {g \over 2}SQ^2.
\end{equation}
Here $Q$ is another chiral superfield and $g$ is
a coupling constant. The perturbative dynamics of this system is not
difficult to understand. For  large values 
\begin{equation}
  |S| >> S_c = {\Lambda \over \sqrt{g}}
\end{equation}
the $Q$-particle
has a zero vacuum expectation value (VEV), a large mass $\sim gS$ and 
it decouples. By integrating
it out we get an effective superpotential given by eq. (2). Moreover  
the one-loop
corrected K\"ahler potential has the form
\begin{equation}
K = SS^+\left ( 1 - {g^2 \over 16\pi^2}{\rm ln}{SS^+ \over M_P^2}  
\right)
\end{equation}
where  our ultraviolet cut-off is the reduced Planck mass $M_P$.
This simply translates  into
the following effective potential \cite{dss}
\begin{equation}
 V(S)= \Lambda^4 \left ( 1 + {g^2 \over 16\pi^2}ln {SS^+ \over M_P^2}  
\right )
\end{equation}
which drives inflation. When $S$ drops below $S_c$,  $Q$ picks up a  
nonzero
VEV and cancels the cosmological constant. Note that the end of  
inflation
does not necessarily coincide with this transition, it generically
happens earlier, when the slow-roll
conditions break down.
The scenario based on the above potential was already considered in  
\cite{dss}.
The main drawback of this scenario is that the scale $\Lambda$ is an  
input of the theory. The purpose of this letter is to show that:
$(1)$ inflation with a linear superpotential can be  
induced dynamically; $(2)$
the dynamically generated scale, as suggested by ${\delta \rho \over  
\rho}$ 
can be   identified with the scale of grand unification.
We construct a simple example which achieves these results.

Let us consider an  $SU(2)$ gauge theory with  matter  consisting
of four doublet  chiral superfields $Q_I, \bar Q^J$ ($I,J = 1,2$ are
flavour indices)\footnote{We write two of the doublets as
anti-doublets and distinguish between meson and baryon bilinears  
in eq. (8).
Our notation allows for a straightforward
generalization to $SU(N)$ gauge theory  with $N$ flavors. }.
We also add a singlet superfield $S$ and assume the following
classical (bare) superpotential
\begin{equation}
W_{cl} = g S(Q_1\bar Q_1+Q_2\bar Q_2)
\end{equation}
where $g$ is a Yukawa coupling constant. A similar model was considered in
Refs. \cite{it},\cite{iy} in the context of dynamical supersymmetry breaking.
At the classical level, in the absence of this superpotential  
($g=0$), the 
space of vacua ($D$-flat directions) is parametrized by a set
of complex fields consisting of
$S$ plus the following  6 $SU(2)$ invariants
 (mesons and baryons)
\begin{equation}
  M_I^J = Q_I\bar Q^J, ~~ B = \epsilon^{IJ}Q_IQ_J, ~~ \bar B =  
\epsilon_{IJ}
\bar Q^I\bar Q^J.
\end{equation}
The invariants are however subject to the constraint
\begin{equation}
 \det M - \bar BB = 0
\label{class}
\end{equation}
so that in the end the space of vacua at $g=0$ has complex dimension  
6. In the
presence of the superpotential, the classical moduli space has two  
branches 
described as follows:

1) $S\not = 0$, with $M_I^J=B=\bar B=0$. On this branch the quarks
 get a mass $\sim g S$ from the superpotential and the gauge symmetry
is unbroken. This branch has (complex) dimension 1.

2) $S=0$, with  non-zero mesons and baryons satisfying two  
constraints. One 
is eq. (\ref{class}) while the other is  $F_S={\rm Tr} M=0$. Here the  
gauge  group is broken. The dimension of this branch is 4.

This moduli space is further reduced by quantum effects. In  
particular
a non-zero vacuum energy is generated along  the $S\not = 0$ branch.
This energy provides the cosmological constant that drives inflation
and $S$ plays the role of the inflaton. This is easily established by  
considering
the effective theory far away along $S\not = 0$.
Here the quark fields get masses of order $S$ and decouple.
The effective theory consists of the (free) singlet $S$ plus a pure 
$SU(2)$ gauge sector.
The effective scale $\Lambda_L$ of the low-energy
$SU(2)$ along this trajectory is given to all orders by the 1-loop  
matching
 of the gauge couplings  at the quarks' mass $gS$ and 
reads\cite{russians}-\cite{seibergone}
\begin{equation}
\Lambda_L^6 =  g^2S^2\Lambda^4
\end{equation}
where $\Lambda$ is the scale of the original theory with massless  
quarks.
In the pure $SU(2)$ gauge theory gauginos condense  
\cite{gluino1,gluino2} 
and an effective superpotential $\sim \Lambda_L^3$ is generated
\begin{equation}
W_{eff} = g S \Lambda^2.
\label{linear}
\end{equation}
Thus
\begin{equation}
 F_S = g\Lambda^2 \label{Fs}
\end{equation}
and supersymmetry is broken, with a vacuum energy density $F_S^2$  
which
is independent of $S$: this is a perfect condition for  
inflation\footnote{Different scenarios based on the idea of
dynamical inflation were considered in Refs. \cite{scott,riotto}.}. 
We will later discuss the effects that generate a small
curvature on this flat potential. Notice that
this result for $W_{eff}$ can also be obtained by considering the
confined superpotential for mesons and baryons.
Quantum corrections\cite{quantumc} 
modify eq. (\ref{class}) to
\begin{equation}
 \det M - \bar BB - \Lambda^4 = 0 
\label{quantum}
\end{equation}
Then, with the aid of  a  Lagrange
multiplier field $A$, we can write the full quantum superpotential
for the confined mesons and baryons  as \footnote{See also \cite{masiero}
where a somewhat similar effective $W$ was studied for constant $S$ field.} 
\begin{equation}
 W = A( \det M  - \bar BB - \Lambda^4) + g S {\rm Tr } M.
\end{equation}
This equation shows that at $S\not = 0$ F-flatness cannot be  
satisfied,
so that branch 1) is lifted. It also shows that branch 2) is modified  
by
eq. (\ref{quantum}) but unlifted. Again at $gS\gg \Lambda$ one can
integrate out the composites and obtain the effective result eq.  
(\ref{linear}).



One could also recover this result by an explicit analysis of the  
$F$-terms
behavior for large values of $S$. These F-terms are
\begin{eqnarray}
F_A = {\rm det}M - B\bar B  - \Lambda^4,~~~F_{M_I^J} =  
A\epsilon_{IK}\epsilon^{JL}M_L^K + g\delta^J_I S,\nonumber\\
F_S = g{\rm Tr M}= 0,~~~ F_B = A\bar B, ~~~F_{\bar B} = AB
\end{eqnarray}
For $S\to \infty$, we want to identify the energetically most  
favorable 
path, the one along which the tree-level scalar potential
assumes an (asymptotically) constant value. Such trajectories can be  
classified
by two possible asymptotics for  $F_S$ as $S\rightarrow \infty$:
either $F_S = 0$ or $F_S = const \neq 0$.
First, it is easy to understand that $F_S=0$
gives that either $F_{M_1^1}$ or $F_{M_2^2}$ should grow
with $S$ as $S \rightarrow \infty$. Thus we are lead to a solution
for constant, but nonzero $F_S$. There exists  such a solution with  
all  
other $F$-term vanishing
\begin{equation}
\bar B = B = 0,~~~
M_I^J = \delta ^J_I\Lambda^2,~~~A\Lambda^2/g = -S,~~~
F_S=g\Lambda^2
\label{scaling}
\end{equation}
This matches the result of the gaugino condensation argument.
This is energetically the most attractive trajectory in  field space
subject to $S \rightarrow \infty$.

Again the one-loop corrections to the K\"ahler potential
\begin{equation}
 K = SS^+ \left ( 1 - {g^2 \over  4\pi^2}{\rm ln }(SS^+/M_P^2)\right )
\end{equation}
provide an effective potential via eq. (\ref{linear})
\begin{equation}
V(S) = {|\partial_S W|^2\over \partial^2_{SS^+}K}\simeq  
g^2\Lambda^4\left 
( 1 + {g^2 \over 4\pi^2}{\rm ln}(SS^+/M_P^2)  
\right).  \label{vvv}
\end{equation}
This logarithmic slope pushes $S$ towards the   origin and, for
sufficiently large initial $S >> S_c$, leads to inflation. For large values  
of $S$ the potential is very flat and   gradually becomes steeper as the  
inflaton rolls towards the origin. The slow roll regime is characterized by  
two conditions
\begin{equation}
\epsilon = 2 \left |{M_PV' \over V }\right|^2 << 1,~~~
|\eta| = \left |{M_P^2V'' \over V} \right |<< 1  \label{33}
\end{equation}
where $V$ is the scalar potential, the primes denote  derivatives
with respect to the inflaton field and again $M_P$ is the reduced Planck  
mass.
This fixes a lower bound on the value of $S$ at the end of inflation
\begin{equation}
    S = {g \over 2\pi}M_P
\end{equation}
as a result, for $g < 1$, the last $60$ e-foldings
can take place for values of $S$ well below the Planck scale. This
fact provides a big advantage for avoiding a large gravity-mediated
curvature of the inflaton potential, which is a grave difficulty for
most inflationary scenarios in  supergravity. We will
come back on this point later on.
To be precise, neglecting supergravity corrections, the value of $S$  
at   $N$ e-foldings before the end of inflation is given by \cite{dss}
\begin{equation}
S_N = {g \over 2\pi}\sqrt{2N}M_P.
\end{equation}
The predicted cosmological parameters such as the spectral index
$n = 1 + 2 \eta$, density perturbations and  relative contribution of 
gravitational waves
$R = 12\epsilon$ has to be evaluated at the value of $S$ for which  
the scale of
interest crossed out of the de Sitter horizon.
Because of the logarithmic shape of the potential,
$n$ is independent on either $g$ or $\Lambda$. Indeed $n$ is also 
independent on the gauge group.
We have \cite{dss}
\begin{equation}
n \simeq 1 - {1 \over N},~~~R \simeq {6g^2 \over N\pi^2},~~~
{\delta T \over T} \simeq {1 \over 2} \sqrt{{N\over 45} }\left (  
{\Lambda^2 \over  M_P^2} \right)
\end{equation}
The COBE normalization fixes the scale $\Lambda \sim 10^{16}-  
10^{15}$ GeV,
which is very close to the GUT scale.
This remarkable coincidence suggests the intriguing possibility that,  
in a suitably modified scenario, $\Lambda$ may be the dynamically  
generated grand unification scale\footnote{
See  also Ref. \cite{cheng}, which discusses, from a different perspective,
 a model with a dynamical GUT scale.}. The most 
straightforward way to implement this is to add
a coupling to  matter in the adjoint of the GUT group. Let us  
consider, for
definiteness, $SU(5)$ with an adjoint $\Sigma$ and with  
superpotential
\begin{equation}
W_{tree} = g S(Q_1\bar Q_1+Q_2\bar Q_2)+ {g'\over 2}S{\rm Tr  
}\Sigma^2 + 
{h \over 3}{\rm Tr} \Sigma^3.
\label{guttree}
\end{equation}
Remarkably, this theory has, at the classical
level, the same moduli space of the original one (without $\Sigma$).  
This
was characterized in the points 1) and 2) above. It is easily
checked. Along $S\not =0$, the $F$-flatness
conditions for the $Q$'s and $\bar Q$'s saturated to form invariants  
give
$SM_I^J=SB=S\bar B=0 \to M=B=\bar B=0$. Then $F_S=\tr \Sigma^2=0$  
implies  
$\Sigma=0$, so that the branch $S\not = 0$ is the same as before. At  
$S=0$,
we have $F_\Sigma =0 \to h \Sigma^2=0$ so that also on this branch  
$SU(5)$ is 
unbroken and the moduli space is described by 2) above. In conclusion,
$SU(5)$ is unbroken all over the classical moduli space, and along
the $S\not =0$ branch all other chiral superfields are massive  
$m_{Q\bar Q}=
g S$ and $m_\Sigma\sim g' S$. Then, as long as $SU(5)$ is weak
or in a phase where it cannot generate a superpotential, the quantum  
dynamics
far away along $S$ is the same as before: $W_{eff}= g S\Lambda^2$ and
K\"ahler logs drive inflation. Now there is also a contribution  
$\delta K
\sim g'^{2}\ln SS^+$, but it  does not change the qualitative  
conclusions
and it is even quantitatively negligible for $g'<g$ (which we will
later argue to be favored by other considerations). Thus
also the modified theory in eq. (\ref{guttree}) naturally produces  
inflation.
The interesting differences with respect to the previous case arise
when considering the full quantum moduli space described by
\beq
W_{eff}=A\left (\det M-B\bar B-\Lambda^4\right )+ S\left (g\tr  
M+{g'\over 2}\tr
\Sigma^2\right )+{h\over 3}\tr \Sigma^3.
\label{gutquantum}
\eeq
The $S=0$ branch goes through unchanged, as here $\Sigma=0$. At  
$S\not =0$ there
however appear two isolated points where $SU(5)$ is Higgsed  
respectively
to $SU(4)\times U(1)$ and to $SU(3)\times SU(2)\times U(1)$. The one
of physical interest is given by
\begin{eqnarray}
\langle \Sigma \rangle = \Lambda\sqrt {{-2g \over 15 g'}}  
(2,2,2,-3,-3),~~~
M_I^J = \delta_I^J\Lambda^2,~~B = \bar B = 0,\nonumber\\
S = \Lambda h \sqrt{{-2g \over 15}}g^{'-{3 \over 2}},~~A =  
\Lambda^{-1}h\sqrt{{2 \over 15}}\left ({-g \over g'}\right )^{{3  
\over  
2}}
\end{eqnarray}
It is interesting that in this theory $SU(5)$ is unbroken on the  
classical
moduli space. Quantum effects introduce a deformation that $a$)  
smoothes one
branch  at $S=0$ and  $b$) creates  $SU(5)$ breaking isolated vacua  
near
the origin. At the end of inflation the system will relax into one of  
its
vacua. At this stage it is difficult to decide whether the vacuum
with $SU(3)\times SU(2)\times U(1)$ residual group will be preferred  
over 
the one with $SU(4)\times U(1)$. It is however possible to argue that
the points with broken $SU(5)$ are preferred over the continuum where
$\Sigma=0$.
As we have seen, inflation  starts out with large and slowly  
decreasing  $S$ and
$A$ (and with ``frozen'' mesons) satisfying eq. (\ref{scaling}). 
Notice that in the supersymmetric isolated minima
with broken $SU(5)$, the fields $B,\bar B, M, S,   A$
satisfy a special case of eq. (16). In other words
the points with broken $SU(5)$ lie on the inflationary branch. 
This suggests that at the end of  inflation the system   
will very likely
first relax to one of these vacua. Moreover, at least for
$g'<g$,  there is further indication that $SU(5)$, and not $SU(2)$,  
ends up being broken. This is beacuse in this limit $\Sigma$ becomes  
tachionic before the other fields, and drives $F_S$ to zero. 
It is clear that there still  remains the question of whether the system may
``overshoot'' into  the other branch of vacua at $S=0$. 
This question, as the one concerning 3-2-1 versus  
4-1 GUT breaking, will not be discussed here. It involves 
control over the behaviour of the K\"ahler potential in
the strong coupling region (small $S$).  This is beyond
the purposes of our paper. Our model is a simple prototype illustrating the
general idea. Nonetheless, 
we find that the feature we can control, the occurrence of an 
inflationary plateau ending in GUT breaking isolated minima, is 
interesting {\it per se}.

 In supergravity the Planck scale suppressed corrections to the  
K\"ahler potential usually generate
a large curvature to the inflaton
potential and are incompatible with the slow-roll
assumption. In our case the situation looks much more promising
due to the fact that: $1)$ The inflaton superpotential is linear and
$2)$ inflation (at least for the last 60 e-foldings) can take place  
for values of $S$ below $M_P$. For generic K\"ahler metric,  
our inflationary potential
can be expanded as
\begin{equation}
V = g^2\Lambda^4 \left [ 1 + {g^2 \over 4\pi}ln{SS^+ \over M_P^2} +
\alpha{SS^+ \over M_P^2} + {\rm higher~powers} \right ]
\end{equation}
Here the bilinear  comes from the non-minimal term
$\alpha {(SS^+)^2 / M_P^2}$ in the K\"ahler potential.
The fact that there is no contribution to the $\alpha$-term from the
canonical part $SS^+$ of the K\"ahler potential is a result of the
effective linear superpotential (13). The same cancellation was
observed in \cite{copeland} in the context of a non-dynamical model.
However the inflaton potential obtained by those authors was very  
different
since the radiative corrections were not considered.
Unless $g$ is tuned to a very small
value these corrections are very important and 
lead to rather different conclusions. The contribution to the 
inflaton mass
from the canonical K\"ahler metric does not vanish
for other type of superpotentials.
This is why the generic potentials, which lead to inflation in the
global SUSY limit, in general have a problem (already in minimal)
supergravity. (The case of
zero effective inflationary superpotential, with $D$-term driven  
inflation
is exceptional and avoids this problem\footnote
{An alternative possibility is
the inflaton being a pseudo-Goldstone mode\cite{gold}.}\cite{dterminf}).
To satisfy the slow roll conditions all we need is that  
$\alpha$ is somewhat smaller than one. Depending on the
balance between the parameters one can face different regimes.
For $g \sim 0.1- 1$, the
end of inflation is practically always triggered by the log-term whenever
inflaton drops
to (20), irrespective of the value of $\alpha$ (provided the slow roll
conditions are not broken by this term at the beginning).
The earlier stages are more sensitive to the precise
values of $g$ and $\alpha$. For $g \lsim 0.1$ the
beginning of last $60$
e-foldings can be dominated by gravity corrections if $\alpha$ is  
large enough\cite{linderiotto}
(crudely, the evolution above $S_N$ will be
dominated by the $\alpha$-term provided
$\alpha > 1/2N$). For $g\gsim 1$ higher supergravity terms,
like $\Lambda^4(SS^+)^2/M_P^4$ in $V(S)$, will 
dominate the early stage of inflation, even for arbitrarily small $\alpha$  
\cite{gr}.
Notice that for $gS \sim \Lambda$ the nonperturbative corrections to  
the K\"ahler potential from
gaugino condensation become ${\cal O}(1)$.
 These corrections switch-off
as $\sim \Lambda^4/(g^2S^{4})$ \cite{us} for large values of $S$ and  
therefore  can dominate
the potential at the later stages of inflation if $g$ is small. In conclusion
 the predictions on the smaller scales (small $N$
region in eqs. (21-22)) can be affected by gaugino condensation 
when $g$ is $\lsim 0.1$, whereas  those at large scale can be modified by
supergravity corrections for $g\gsim 1$.
The predictions for intermediate $N$ are   less 
sensitive to the variation of $g$. The values of $g$ for which
gravity and strong dynamics effects on eq. (22) can be ignored is  
somewhere  between $0.1$ and $1$.

  In conclusion we have constructed a scenario with dynamically  
generated
inflaton potential, and without a slow-roll problem. The scenario is  
quite restrictive and, under reasonable assumptions, the predicted  
cosmological
observables  depend on a single parameter, the dynamically generated scale, 
which is also of the order of the GUT scale. 
This lead us to think that both inflation
in the early Universe and the spontaneous breaking of the GUT  
symmetry  in
the present vacuum may have a  common dynamical origin. We have shown that
within the given inflationary scenario this connection can be
naturally established.

We acknowledge fruitful discussions with A. Brignole, J. Garcia-Bellido,
G. Pollifrone, C. Savoy and G. Veneziano.

\end{document}